# Quantum chemistry studies of the O K-edge X-ray absorption in $WO_3$ and $AWO_3$

Dmitry Bocharov*, Alexei Kuzmin, Juris Purans, and Yuri Zhukovskii
Institute of Solid State Physics, University of Latvia, 8 Kengaraga Str., LV-1063 Riga, Latvia

## ABSTRACT

In this work we present an interpretation of experimental O K-edge x-ray absorption near edge structure (XANES) in perovskite-type $WO_3$ and $AWO_3$ compounds (A = H and Na) using three different first principles approaches: (i) full-multiple-scattering (FMS) formalism (the real-space FEFF code), (ii) hybrid density functional theory (DFT) method with partial incorporation of exact Hartree-Fock exchange using formalism of the linear combination of atomic orbitals (LCAO) as implemented in the CRYSTAL code; (iii) plane-wave DFT method using formalism of the projector-augmented waves (PAW) as implemented in the VASP code.

## 1. INTRODUCTION

Tungsten trioxide $WO_3$ is a well known electrochromic material, which is used in switchable technology for glazing, mirrors, transparent displays and a variety of other applications. The electrochromic effect in $WO_3$ appears as a color change upon intercalation of a small ion A (A = $Li^+$, $H^+$ etc) leading to a lowering of the tungsten ion valence state and a formation of tungsten bronze $A_xW^{(6-x)+}O_3$ [1]. The co-injected electrons are introduced in to the tungsten $5d$ states, located at the bottom of the conduction band, therefore the study of the electronic structure variation in tungsten oxides is of crucial importance for detailed understanding of electrochromism.

The unoccupied states above the Fermi level can be probed by x-ray absorption spectroscopy. By tuning the x-ray energy across the tungsten or oxygen absorption edges, different final states can be studied. In particular, the W $L_3$-edge provides us with the information on the tungsten $d$ states, whereas the O K-edge is sensitive to the oxygen $p$ states. The interpretation of x-ray absorption near edge structure (XANES) is generally a difficult task in spite of the great progress during last ten years [2]. This complexity arises mainly from the necessity to treat the excited states, including the relaxation phenomena, and from an intermediate character of empty electron states located close to the Fermi level. Note that the relaxation effects, due to the interaction between the core-hole and the excited electron, are smaller for the K-edge than for the $L_3$-edge, therefore the K-edge spectra can be interpreted based on the results of the ground state calculations as provided by many quantum chemistry codes.

In this work we will discuss interpretation of the oxygen K-edge XANES signals in perovskite-type $Na_xWO_3$, $H_xWO_3$, $WO_3$ and structurally similar $ReO_3$ and $H_xReO_3$ compounds, which have been experimentally studied previously [3-5]. Note that these XANES signals [3-5] are quite similar in overall shape but differ in intensity and position of some features, in particular, close to the absorption edge (or Fermi level).

Nowadays most common approach to the interpretation of XANES is based on the real-space full-multiple-scattering (FMS) formalism as, for example, realized in the FEFF code [6, 7]. It includes the final state effects of the self-energy and the core-hole, but uses spherical muffin-tin (MT) potentials and self-consistency is realized for the electron density, but not for the total energy of the system. A more sophisticated approach, developed recently by Joly [8], uses a finite cartesian grid in real space, thus overcoming the problem of spherical MT potential, however, the method is currently still not self-consistent.

Further we will apply three different first principles methods, as realized in the FEFF [6, 7], CRYSTAL [9] and VASP [10] codes, to the interpretation of the O K-edge XANES signals. The results obtained by the conventional multiple-scattering approach, realized in the FEFF code, will be compared with those obtained by the quantum chemistry LCAO (localized atomic orbitals) and PAW (projected augmented waves) methods. The charge analysis will be also presented.

---

* bocharov@latnet.lv

## 2. THEORETICAL MODELLING

The x-ray absorption coefficient $\mu(E)$ in the one-electron and dipole ($\Delta l=\pm 1$, $l$ is the orbital momentum) approximations is given by the Fermi's Golden Rule:

$$\mu(E) \propto \left|\langle \phi_f | \hat{\varepsilon} \cdot \vec{r} | \phi_i \rangle\right|^2 \rho(\hbar\omega + E_i - E_f). \tag{1}$$

Here $\phi_i$ and $\phi_f$ are the wave functions of the initial and final states, $\rho$ is the electronic density of unoccupied final states (DOS) located above the Fermi level, $\hbar\omega+E_i \equiv E$ is the photoelectron energy, $E_i$ is the core-level energy. The first multiplier in Eq. (1) is the transition matrix element, being a smooth function without any sudden changes and abrupt peaks. Therefore, all peaks in the absorption spectrum are due to variations in the $\rho(E)$ DOS function, describing relaxed final state in the presence of the core hole at the absorbing atom. In the dipole approximation, one deals at the O K-edge with the excitation of an electron from the 1$s$(O) core-level into the excited states having $p$ character and, therefore, $\rho(E)$ reflects the projected DOS.

Further we will compare three different approaches to the DOS calculation: the conventional method, based on the multiple-scattering formalism and taking into account the relaxation effects, and two quantum chemistry methods dealing with a ground-state DOS.

### 2.1. Full-Multiple-Scattering (FMS) approach.

Full-multiple-scattering (FMS) calculations of the O K-edge XANES signals have been performed using the *ab initio* real-space FEFF 8.2 code within the dipole approximation [6, 7]. A self-consistent FMS calculations have been done with a complex Hedin-Lundqvist exchange-correlation potential within muffin-tin approximation for a cluster [11], representing a fragment of the cubic AWO$_3$ (A = H, Li, Na, K, Rb, Cs) perovskite-type structure, defined by the lattice parameter ($a$) and centered at oxygen atom (absorber). The radii of the clusters have been varied up to 8.0 Å and included up to 13 coordination shells. Besides the XANES signals, the partial DOS have been also calculated for the absorbing oxygen atom O* and all neighboring atoms (O, W, A). Note that the FEFF calculations include explicitly the core-hole relaxation effects, and therefore we have used them to estimate the effect of the core-hole on the spectral shape of the XANES signal.

### 2.2. Hybrid DFT-HF method based on the formalism of linear combination of atomic orbitals (LCAO).

Main algorithms used in the CRYSTAL code employ the Gaussian-type functions centered on atomic nuclei as basis sets (BSs) for an expansion of the crystalline orbitals. The BSs for all atoms were first taken from the website of CRYSTAL code [9] or other libraries of atomic basis sets and then slightly re-optimized by us for each AWO$_3$ compound. The HAYWSC (Hay-Wadt small-core) pseudopotential [12] is used for the sub-valence atomic cores. The effective core pseudopotential for W atom (ECP) was constructed elsewhere [13].

The choice of hybrid exchange-correlation potentials for our calculations is caused by several shortcomings when calculating the electronic structure of both insulators and semiconductors using methods based on formalisms of Density Functional Theory (DFT) and Hartree-Fock (HF), especially when estimating a band gap between the top valence band and the bottom of conduction bands: the HF calculations markedly overestimate this value whereas the DFT calculations usually underestimate it. To avoid this artifact, the CRYSTAL code realizes a partial incorporation of the exact HF exchange into the DFT exchange potential (with varying mixing ratio) as implemented in both B3LYP and B3PW exchange-correlation potentials [9]. The test calculations for our AWO$_3$ systems give a preference to none of methods, however, in our current study we have used predominantly B3PW since for analogous calculations in other perovskites (e.g., SrTiO$_3$ [14]), this potential can give the better results.

In this study, different combinations of atomic BSs and exchange-correlation potentials have been used. For further analysis, we employ various BS combinations as described in the next sections. The given nomenclature is shown in Table 1 where one can see: (i) the labels used in the article, (ii) the standard BS labeling for each atom basis set, (iii) a source (article or data base) which these data are taken from, (iv) a type of exchange-correlation potential.

Basis sets and lattice parameters have been optimized using the ParOptimize code [15] with the requirement of the total energy minimum following two criteria. First, the calculated lattice constants should be close to the experimental values. Second, the bulk modulus should be close to those from other theoretical calculations since no experimental data are available so far. The obtained values of parameters are compared with data from other studies in Tables 3 and 4.

Table 1. Basis sets used in CRYSTAL calculations.

| Compound | Basis label in article | Basis set type, source of basis set* | | exchange-correlation potential |
|---|---|---|---|---|
| | | A ion | O | |
| $WO_3$ | $WO_3$-B3PW-Cora | - | O{8s-511sp}[13] | B3PW |
| | $WO_3$-B3PW-DZ | - | cc-pVDZ [16] | B3PW |
| | $WO_3$-B3PW-631G | - | O{6s-31sp-1d}[9] | B3PW |
| | $WO_3$-B3PW-8831G | - | O{8s-411sp}[9] | B3PW |
| | $WO_3$-B3LYP-Cora | - | O{8s-511sp} | B3LYP |
| $HWO_3$ | $HWO_3$-B3PW-Cora-Dovesi | H{5s-11sp} [9] | O{8s-511sp} | B3PW |
| | $HWO_3$-B3PW-Cora-Towler | H{8s-211sp}[17] | O{8s-511sp} | B3PW |
| $LiWO_3$ | $LiWO_3$-B3PW-Cora-Merawa | Li{5s-11sp} [18] | O{8s-511sp} | B3PW |
| | $LiWO_3$-B3PW-Cora-Ojamae | Li{6s-11sp} [19] | O{8s-511sp} | B3PW |
| $NaWO_3$ | $NaWO_3$-B3PW-Cora | Na{8s-511sp} [9] | O{8s-511sp} | B3PW |
| | $NaWO_3$-B3PW-DZ | Na{8s-511sp} [9] | cc-pVDZ | B3PW |
| | $NaWO_3$-B3LYP-DZ | Na{8s-511sp} [9] | cc-pVDZ | B3LYP |
| $KWO_3$ | $KWO_3$-B3PW-Cora | K{8s-6511sp}[20] | O{8s-511sp} | B3PW |
| $RbWO_3$ | $RbWO_3$-B3PW-Cora | Rb{HAYWSC-31sp} [9] | O{8s-511sp} | B3PW |
| $CsWO_3$ | $CsWO_3$-B3PW-Cora | Cs{HAYWSC-31sp} [9] | O{8s-511sp} | B3PW |

* The effective core pseudopotential for W (ECP) for all calculations was given from [13]

### 2.3. DFT method based on the formalism of plane-wave functions.

As a second *ab initio* quantum chemistry method we have applied the DFT computational scheme as implemented in computer code VASP 4.6 [10] using the formalism of projector-augmented waves (PAW) including the plane-wave basis sets. For the calculations we have chosen the non-local Perdew-Wang-91 exchange-correlation functional [21] and the scalar PAW pseudopotentials representing the core electrons of W, O and Na atoms containing different number of valence electrons. The basis sets, used in the calculations, are given in Table 2. An extension "*h*" (hard) implies that the potential is harder than the standard potential and hence requires a larger energy cut-off, whereas "*s*" (soft) means that the potential is softer than its standard version [10].

As in case of CRYSTAL calculations we compare the calculated bulk modulus and lattice parameters with the corresponding data obtained in other studies (see Table 3 and Table 4).

Table 2. Basis sets used in VASP calculations.

| Compound | Basis label in article | Basis set type, number of valence electrons (v.e.). | | |
|---|---|---|---|---|
| | | A ion | W | O |
| $WO_3$ | $WO_3$-W12-O6s | - | $5p^6 5d^4 6s^2$ (12 v.e.) | $2s^2 2p^4\_s$ (6 v.e.) |
| | $WO_3$-W12-O6h | - | $5p^6 5d^4 6s^2$ (12 v.e.) | $2s^2 2p^4\_h$ (6 v.e.) |
| | $WO_3$-W6 -O6s | - | $5d^4 6s^2$ (6 v.e.) | $2s^2 2p^4\_s$ (6 v.e.) |
| $NaWO_3$ | $NaWO_3$-Na1-W6-O6s | $3s^1$ (1 v.e.) | $5d^4 6s^2$ (6 v.e.) | $2s^2 2p^4\_s$ (6 v.e.) |
| | $NaWO_3$-Na9-W12-O6h | $2s^2 2p^6 3s^1$ (9 v.e.) | $5p^6 5d^4 6s^2$ (12 v.e.) | $2s^2 2p^4\_h$ (6 v.e.) |
| | $NaWO_3$-Na9-W12-O6s | $2s^2 2p^6 3s^1$ (9 v.e.) | $5p^6 5d^4 6s^2$ (12 v.e.) | $2s^2 2p^4\_s$ (6 v.e.) |

# 3. RESULTS AND DISCUSSION

The experimental O K-edge XANES signals in $WO_3$, $HWO_3$ and $Na_xWO_3$ ($x$ changes from 0 to 1) have been reported and discussed in details previously [3-5]. They show several tendencies upon hydrogen or sodium incorporation in tungsten trioxide, among which an increase of the first peak intensity is the most prominent one. Since the origin of the first peak is related to the unoccupied states at the bottom of the conduction band, one can expect that the quantum chemistry methods using more refined self-consistent procedures should provide with more accurate estimates for DOS variations than the multiple-scattering approach using the muffin-tin potential approximation.

The O K-edge XANES signals, calculated by the FEFF code, are shown in Figure 1. They are in qualitative agreement with the available experimental data [4, 5] and show more resolved fine structure due to the smaller broadening effects. However, there is some disagreement between the energy scale of the calculated and experimental curves attributed to the inaccuracies in the energy-dependent exchange-correlation potential. Note that the calculated XANES spectra in the edge region depend strongly on the assigned Fermi level position. In this study, the position of the Fermi level was shifted in all cases by 2 eV relative to the value obtained during the self-consistent procedure, to achieve the best agreement with the experimental XANES signals. The first peak intensity depends also on the A ion type but this effect correlates strongly with the position of the Fermi level through a variation of the respective states occupancy number. The FEFF code accounts for the presence of the core hole at the absorbing atom and, thus, provides with both relaxed $p(O^*)$ and unrelaxed $p(O)$ oxygen partial DOS: close agreement between the two curves suggests that a ground state calculations can be used in this case. This conclusion makes strong support for the use of quantum chemistry calculations for interpretation of our XANES spectra.

Thus, we will discuss further the results by CRYSTAL and VASP codes. The influence of the oxygen basis sets on the projected DOS (PDOS) in $WO_3$ is shown in Figure 2. Note that the W and O basis functions were optimized in all calculations. As one can see, the obtained PDOS are in agreement only for the energies up to 15 eV above the Fermi level. For higher energies, the LCAO method works rather badly due to the basis-set limitations.

Theoretically calculated lattice constants are compared with the experimental values in Table 3. They vary within the narrow limit of about ±0.05 Å and are in agreement with the experimental data within the accuracy of about ±0.15 Å. Note that the lattice constant of $LiWO_3$ in two different experimental studies differ by 0.16 Å [22, 23]. The calculated bulk moduli are given in Table 4 and are close to those from other theoretical calculations [25, 26]. The values, obtained by CRYSTAL and VASP codes, are within the limits of 20% for $WO_3$ and 10% for $NaWO_3$.

The covalency of bonds (Tables 5 and 6) have been evaluated using both Mulliken and Bader charge analyses when using the CRYSTAL and VASP codes, respectively. The W-O bond has a middle-strong covalence while O-A bond is more ionic. The O-H bond is the only exception from this rule for the $A^+$ ions and shows a strong covalency. Note that the covalency of the W-O bond is the main responsible for the intensity of the first peak, located at about 3 eV above the Fermi level in the O K-edge XANES signals (Figure 1), because it reduces the occupation number of the oxygen $2p$ states due to the back-charge transfer to tungsten ions.

The different PDOS for O, W and A ions calculated by both CRYSTAL and VASP codes are compared with the experimental XANES signals in Figures 3 and 4. The theoretical calculations suggest that the first peak arises owing to the $p(O)$ and $d(W)$ hybridized states. Both calculations show that the $s(Na)$ states are located at ~5 eV above the Fermi level. It can promote an increase of the second peak intensity, visible in the experimental spectra. CRYSTAL calculation show that $s(H)$ states give contribution close to the Fermi level, therefore it can be responsible for an increase of the first peak upon hydrogen intercalation into tungsten oxide.

Note that both CRYSTAL and VASP codes give comparable accuracy for the energy states close to the Fermi level. However, the PAW method, implemented in VASP, leads to better results at higher energies (greater than 15 eV above the Fermi level) due to a localization of the LCAO basis used in the CRYSTAL code.

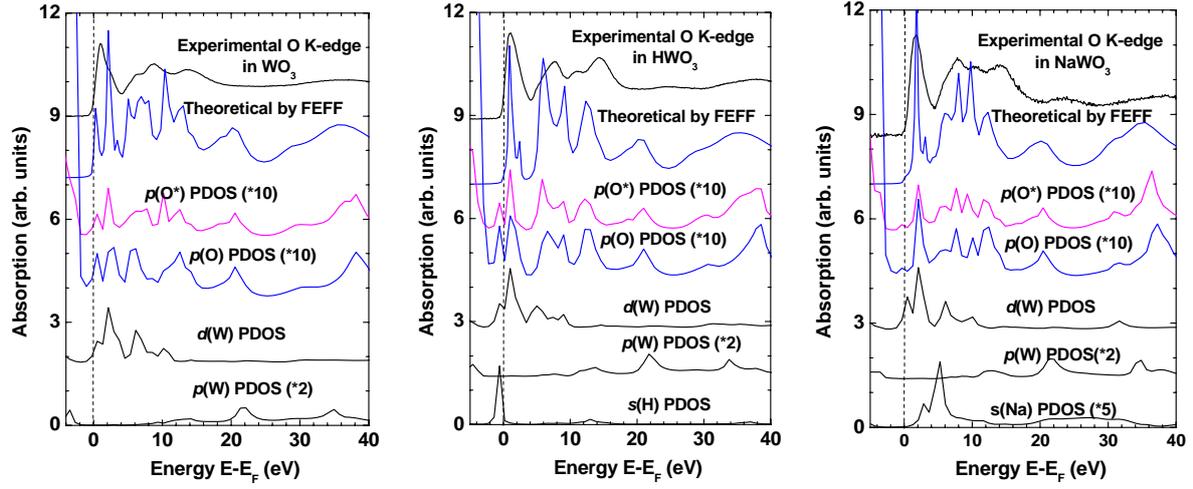

Figure 1: Theoretical O K-edge XANES in $WO_3$, $HWO_3$ and $NaWO_3$, calculated by the FEFF code, for the cluster, having the radius $R = 7.84$ Å and the lattice constant a = 3.86 Å. The position of the Fermi level ($E_F$) was shifted by 2 eV in all cases. PDOS for $s$(Na), $p$(W), $d$(W), $p$(O) un $p$(O*) are also shown (O* is the absorbing atom). DOS amplitude was multiplied by the reported factors. Spectra are vertically shifted for clarity.

Table 3. Theoretically calculated and experimental values of the lattice constants.

| Compound | Theoretically calculated lattice constants in this work, Å | | Experimentally obtained lattice constants, Å |
|---|---|---|---|
| | CRYSTAL | VASP | |
| $WO_3$ | 3.775 ($WO_3$-B3PW-Cora), 3.806 ($WO_3$-B3PW-DZ), 3.791 ($WO_3$-B3PW-631G), 3.759 ($WO_3$-B3PW-8831G), 3.787 ($WO_3$-B3LYP-Cora) | 3.84 ($WO_3$-W12-O6s) 3.839 ($WO_3$-W12-O6h) 3.828 ($WO_3$-W6 -O6s) | 3.8144 [22], 3.78 [23] |
| $HWO_3$ | 3.788 ($HWO_3$-B3PW-Cora-Dovesi), 3.834 ($HWO_3$-B3PW-Cora-Towler) | | 3.8235 [22], 3.78 [23] |
| $LiWO_3$ | 3.837 ($LiWO_3$-B3PW-Cora-Merawa), 3.845 ($LiWO_3$-B3PW-Cora-Ojamae) | | 3.8712 [22], 3.71 [23] |
| $NaWO_3$ | 3.865 ($NaWO_3$-B3PW-Cora), 3.883 ($NaWO_3$-B3PW-DZ), 3.899 ($NaWO_3$-B3LYP-DZ) | 3.91 ($NaWO_3$-Na1-W6-O6s), 3.92 ($NaWO_3$-Na9-W12-O6h), 3.921 ($NaWO_3$-Na9-W12-O6s) | 3.8911 [22], 3.87 [23], 3.85 [24] |
| $KWO_3$ | 3.927 ($KWO_3$-B3PW-Cora) | | 3.9375 [23] |
| $RbWO_3$ | 4.016 ($RbWO_3$-B3PW-Cora) | | 3.9818 [23] |
| $CsWO_3$ | 4.159 ($CsWO_3$-B3PW-Cora) | | 4.0409 [23] |

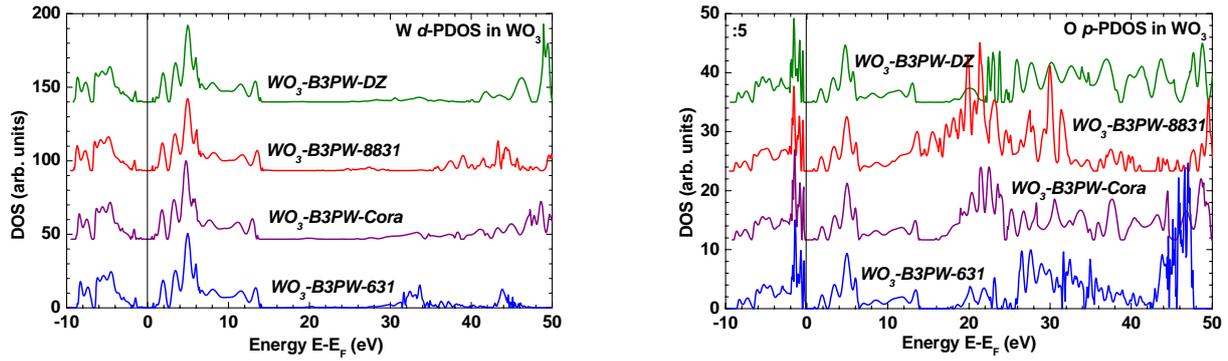

Figure 2: PDOS for $d$(W) and $p$(O) in $WO_3$ calculated by CRYSTAL code using different oxygen basis sets. Spectra are vertically shifted for clarity. The spectra in a zone between -10 and 0 eV on $p$(O) graph has been divided by 5. The energy scale is relative to the Fermi energy ($E_F$).

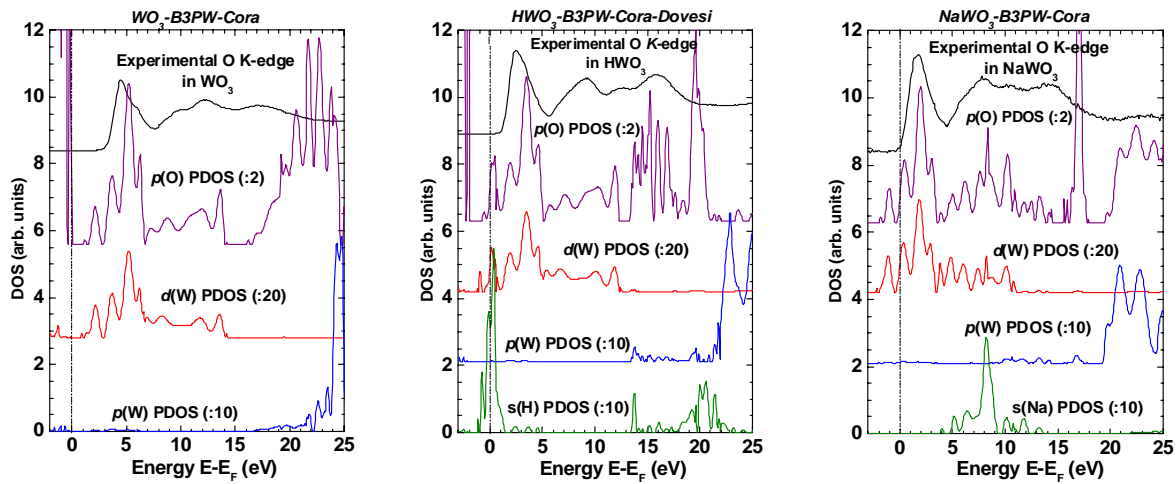

Figure 3: Theoretical O K-edge XANES in $WO_3$, $HWO_3$ and $NaWO_3$ calculated using CRYSTAL code. DOS amplitude was multiplied by the reported factors. Spectra are vertically shifted for clarity. The energy scale is relative to the Fermi energy ($E_F$).

Table 4. Theoretically calculated bulk moduli in comparison with the data from other theoretical works

| Compound | Theoretically calculated bulk modulus in this work | | Theoretically calculated bulk modulus from other works |
|---|---|---|---|
| | CRYSTAL | VASP | |
| $WO_3$ | 242 ($WO_3$-B3PW-Cora), 220 ($WO_3$-B3PW-DZ), 226 ($WO_3$-B3PW-631G), 258 ($WO_3$-B3PW-8831G), 234 ($WO_3$-B3LYP-Cora) | 249 ($WO_3$-W12-O6s), 237 ($WO_3$-W12-O6h), 233 ($WO_3$-W6 -O6s) | 224 (FP-LMTO, GGA) [25], 256 (FP-LMTO, LDA) [25], 254 (FP-LMTO, LDA) [26], 257 (HF) [26], 281 (HF+correlation) [26] |
| $NaWO_3$ | 216 ($NaWO_3$-B3PW-Cora), 199 ($NaWO_3$-B3PW-DZ), 193 ($NaWO_3$-B3LYP-DZ) | 214 ($NaWO_3$-Na1-W6-O6s), 213 ($NaWO_3$-Na9-W12-O6h), 212 ($NaWO_3$-Na9-W12-O6s) | 203 (FP-LMTO, GGA) [25], 255 (FP-LMTO, LDA) [25], 241 (FP-LMTO, LDA) [26], 233 (HF) [26], 263 (HF+correlation) [26] |

FP-LMTO – Full Potential Linear muffin-tin orbitals,
LDA – local density approximation,
GGA – generalized gradient approximation.

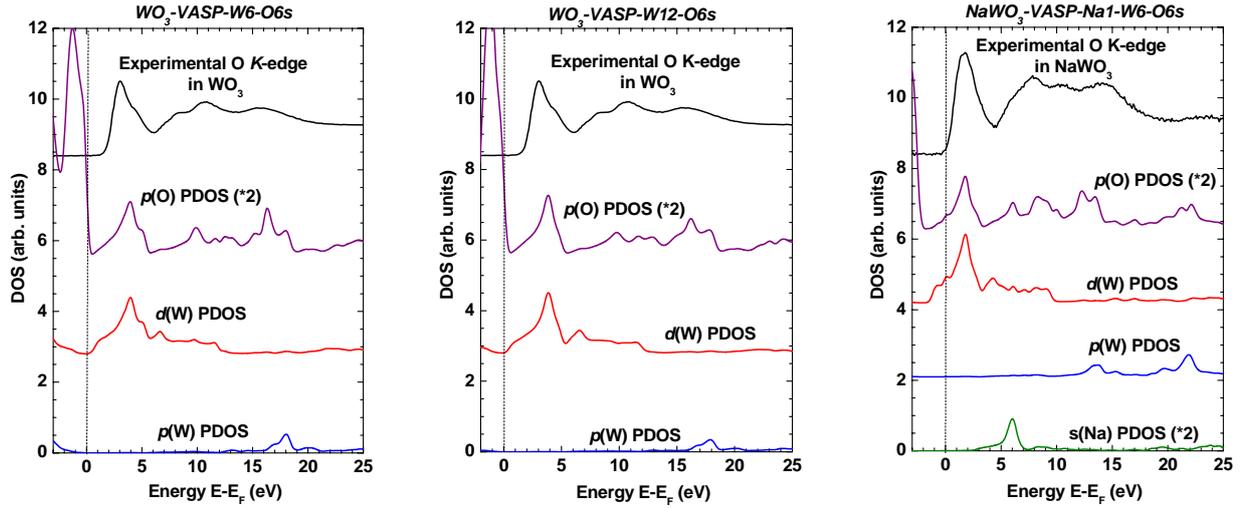

Figure 4: Theoretical O K-edge XANES in $WO_3$ and $NaWO_3$, calculated using VASP code. DOS amplitude was multiplied by the reported factors. Spectra are vertically shifted for clarity.

Table 5. Ion charges using Bader analysis (calculations by VASP).

| Compound | Potential | Ion charge ($e$) | | |
|---|---|---|---|---|
| | | A ion | W | O |
| $WO_3$ | $WO_3$-W6-O6s | - | +3.0536 | -1.0178 |
| | $WO_3$-W12-O6s | - | +3.1147 | -1.0382 |
| | $WO_3$-W12-O6h | - | +2.9997 | -0.9999 |
| $NaWO_3$ | $NaWO_3$-Na1-W6-O6s | +1 | +3.61 | -1.54 |
| | $NaWO_3$-Na9-W12-O6s | +0.8984 | +2.7738 | -1.2241 |
| | $NaWO_3$-Na9-W12-O6h | +0.899 | +2.7808 | -1.2266 |

Table 6. Ion charges using Mulliken analysis (calculations by CRYSTAL).

| Compound | Potential | Ion charge (e) | | |
|---|---|---|---|---|
| | | A ion | W | O |
| $WO_3$ | $WO_3$-B3PW-Cora | - | +2.917 | -0.972 |
| | $WO_3$-B3PW-DZ | - | +2.983 | -0.994 |
| | $WO_3$-B3PW-631G | - | +3.05 | -1.017 |
| | $WO_3$-B3PW-8831G | - | +3.121 | -1.040 |
| | $WO_3$-B3LYP-Cora | - | +2.984 | -0.995 |
| $HWO_3$ | $HWO_3$-B3PW-Cora-Dovesi | +0.009 | +2.686 | -0.898 |
| | $HWO_3$-B3PW-Cora-Towler | -0.121 | +2.78 | -0.886 |
| $LiWO_3$ | $LiWO_3$-B3PW-Cora-Merawa | +0.957 | +2.649 | -1.202 |
| | $LiWO_3$-B3PW-Cora-Ojamae | +0.565 | +2.472 | -1.013 |
| $NaWO_3$ | $NaWO_3$-B3PW-Cora | +0.855 | +2.6 | -1.152 |
| | $NaWO_3$-B3PW-DZ | +0.594 | +2.608 | -1.068 |
| | $NaWO_3$-B3LYP-DZ | +0.592 | +2.64 | -1.077 |
| $KWO_3$ | $KWO_3$-B3PW-Cora | +1.117 | +2.697 | -1.271 |
| $CsWO_3$ | $CsWO_3$-B3PW-Cora | +1.019 | +2.438 | -1.152 |
| $RbWO_3$ | $RbWO_3$-B3PW-Cora | +1.007 | +2.341 | -1.116 |

## 4. CONCLUSIONS

The O K-edge XANES spectra in $WO_3$, $HWO_3$ and $NaWO_3$ have been interpreted using three different first principles approaches based on the formalisms of full-multiple-scattering and quantum chemistry. All methods give qualitative agreement with the experimental signals allowing us to explain the origin of peaks in the experimental data. However, the present accuracy of theoretical calculations is not sufficient to attain quantitative agreement and, thus, to describe uniquely the observed variations of the first peak in the experimental XANES spectra.

The DOS obtained in CRYSTAL calculations with different oxygen basis sets are similar in a region of about 15 eV above the Fermi level, but differ considerably at higher energies. This fact indicates that the LCAO method is not applicable for energies far above the Fermi level due to the localized nature of basis set functions. At the same time, the use of the PAW method leads to quite satisfactory results at high energies. Both quantum chemistry methods allow us to treat more accurate the states close to the Fermi level than the full-multiple-scattering approach.

## ACKNOWLEDGMENTS

The authors would like to thank A. Gulans, V. Kashcheyevs, S. Piskunov, G. Zvejnieks and J. Chepkasova for fruitful discussions. This research was partly supported by the Latvian Government Research Grants No. 05.0005, 05.1714, 05.1717 and Latvian National research programme in Materials Science. D.B. gratefully acknowledges funding from the European Social Fund (ESF).